\documentclass[aps,twocolumn,prb]{revtex4}
\pdfoutput=1
\makeatletter
\def\@dotsep{4.5}
\makeatother

\usepackage{graphicx}
\usepackage{amssymb}
\usepackage{natbib}
\newcommand{\comment}[1]{}

\begin{document}

\title{Asperity contacts at the nanoscale: comparison of Ru and Au}
\author{Andrea Fortini$^{1}$}
\email{fortini@yu.edu}
\author{Mikhail I. Mendelev$^{2}$}
\author{Sergey Buldyrev$^{1}$} 
\author{David Srolovitz$^{1}$}
\affiliation{ {\rm(1)} Department of Physics, Yeshiva University, 500 West 185th Street, New York, NY 10033, USA \\
{\rm(2)} Materials and Engineering Physics, Ames Laboratory, Ames, IA 50011, USA  }

\begin{abstract}
We develop and validate an interatomic potential for ruthenium based on the embedded atom method framework with the Finnis/Sinclair representation.  We confirm that the new potential yields a stable hcp lattice with reasonable lattice and elastic constants and surface and stacking fault energies.   We employ molecular dynamics simulations to bring two surfaces together;  one flat and the other with a single asperity.  We compare the process of asperity contact formation and breaking in Au and Ru, two materials currently in use in micro electro mechanical system switches.  While Au is very ductile at 150 and 300 K, Ru shows considerably less plasticity at 300 and 600 K (approximately the same homologous temperature).  In Au, the asperity necks down to a single atom thick bridge at separation.  While similar necking occurs in Ru at 600 K, it is much more limited than in Au.  On the other hand, at 300 K, Ru breaks by a much more brittle process of fracture/decohesion with limited plastic deformation.
\end{abstract}

\maketitle

\section{Introduction}
Contacts between surfaces are central to a wide range of technologies and present an interesting array of  fundamental issues.   Contact behavior may involve phenomena as diverse as adhesion, friction, bonding, wear and fracture. Most contact problems are intrinsically multiscale;  involving atomic bonding, nano-scale asperities, defect nucleation, elastic and plastic deformation, surface roughness, far-field loading, etc. The atomistic or nano-scale regime is particularly relevant since this is the length scale at which the first contact between rough surfaces occurs. Furthermore, recent technological applications like Micro Electro Mechanical Systems~\cite{Rebeiz:2003} (MEMS) use mechanical contacts that are only a few micrometers large, such that  nanoscale contact physics plays an even greater role.  

Because of its large electrical conductivity, gold has become a popular material for electrical contacts. Nevertheless, its application to micro-contact technology like MEMS switches has been hindered by a lack of reliability. Extensive surface damage leads to failure of some gold contact devices after only several million open/close cycles.~\cite{Chen:2007fj} The resulting surface damage has been studied experimentally with atomic force microscopes in order to understand the effect of adhesion, thermal dissipation and contamination.~\cite{Hyman:1999,Erts:2002,TORMOEN:2004}  Alloying Au with other metals results in increased hardness, but also in increased resistivity.~\cite{Lee:2006}  Atomic-level simulations~\cite{Cha2004,Song2006,Song2007,Song2007a,Sorensen:1998} and experimental observations\cite{Kuipers:1993,Kizuka:1998} have shown that  the separation of gold contacts is accompanied by substantial plastic deformation, leading to ductile material separation with considerable material  transfer from one side of the contact to the other.  This results in significant contact surface morphology evolution in each cycle. Other pure metals, like ruthenium, have recently been used to build metal contacts in MEMS switches. Ruthenium contacts have proven to be more reliable, routinely surviving millions of cycles without significant degradation of the contacts.~\cite{Chen:2007fj}  While Au contacts have been the focus of several atomistic simulations, to our knowledge, there have been no atomistic simulation studies of contact between Ru surfaces.  Therefore, we do not know whether the fundamental contact evolution mechanisms observed in Au occur for other metals generally, and to the important case of Ru, in particular.  Information of this type can help guide the development of future generations of MEMS contact switches. 

A comparison of Au and Ru contact behavior is also interesting because the crystal structures of Ru (hexagonal closed packed) and Au (face centered cubic) differ: this can lead to fundamental differences in crystal plasticity.  Molecular dynamics (MD) simulation is a powerful tool for studying adhesion, defect formation and deformation on the nano-scale level (that of asperities in MEMS contacts). The first step in performing large scale molecular dynamics simulation of Ru is to establish an appropriate interatomic potential.  We have investigated several potentials for Ru available from the literature, but encountered problems with each of sufficient magnitude to make them inappropriate for contact simulations.  In this article, we develop a new embedded atom method~\cite{Daw:1983fk,Daw:1984uq} (EAM) type potential for Ru. We implement this potential in our molecular dynamics simulations of asperity contact and separation.  The main focus of this paper is a comparison of how asperity contacts form and separate in gold and in ruthenium. 

\section{Simulation method}

\begin{figure}[htbp]
   \centering
   \includegraphics[width=7cm]{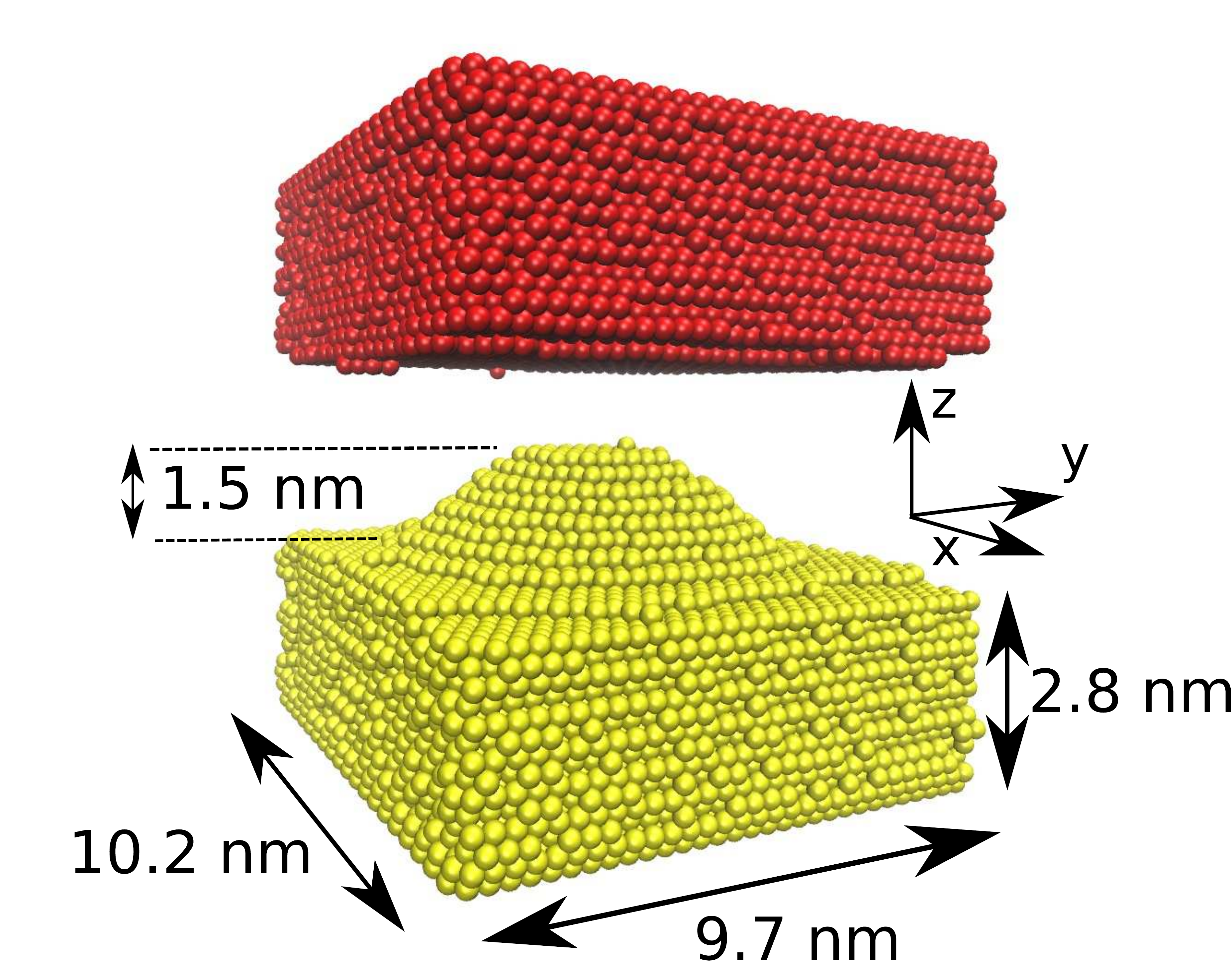}
   \caption{(Color Online) Atomic configuration of the simulation cell after annealing of the asperity to a final temperature T=300 K and prior to the contact simulation experiment. The top and bottom substrates consist of the same material but different colors are used to emphasize material transfer after contact separation.}
   \label{fig:cell}
\end{figure}

In order to understand the formation and separation of nanoscale asperity contacts, we employ the molecular dynamics (MD) technique~\cite{Frenkel2002,Plimpton:1995om} to simulate the accompanying atom-scale dynamics. In MD, the motion of the atoms are simulated by solving Newton's equations of motion for each atom in the simulation cell.  The molecular dynamics were carried out at a constant pressure and temperature.  This was done using the Nos\'{e}-Hoover thermostat~\cite{Hoover:1985yq} and barostat~\cite{Hoover:1986vn} with a time step of $\delta t$=0.0025~ps.  
In the contact experiments, we simulate two substrates, of roughly 15 close-packed planes, facing one another.  These correspond to the (0001) planes of the Ru hexagonal close packed (hcp) lattice and to (111) planes of the Au face centered cubic (fcc) lattice.  On one of the substrates we place an asperity, as shown in Fig.~\ref{fig:cell}. The top two layers in the upper substrate and the bottom two in the lower substrate were kept rigid in order to prevent the two substrates from rotating and to fix the relative displacement of the two substrates in the $z$-direction.  Periodic boundary conditions are enforced in the $x$ and $y$-directions. The simulation cell contains ~35,000 atoms. In the present simulations, an anisotropic isobaric ensemble is used at zero lateral pressure to maintain constant (zero) stress in the $x$ and $y$ directions.  These boundary conditions were chosen to model a periodic surface while allowing the system to expand or contract in the $x$ and $y$-directions, as it would if the substrates were of finite extent in these directions.

Our first step in preparing the system is to construct a cube-shaped asperity with 8 close packed layers on the upper surface of the bottom substrate. 
Second, we anneal the asperity by increasing the temperature from 3 K to 1700 K in 100 ps. Next, we quench the system back to 3 K within 100 ps, and finally we increase the temperature to 600 K or 300K in 100 ps, for Ru and Au, respectively. The system is then quenched to the target temperature that may differ from the above values. This approach allows the asperity shape to evolve towards one that is relaxed away from its initial cubical shape.  The resultant Ru asperity at 300 K is shown in Fig.~\ref{fig:cell}.  

After the asperity annealing is completed, we displace the upper substrate toward the lower substrate at a constant velocity of 0.07 ${\rm \AA/ ps}$ (7 m/s), while holding the lower substrate fixed (i.e., we hold the two atomic layers at the bottom of the lower substrate fixed).  When the distance between the top of the upper substrate and the bottom of the lower substrate is 1.37 nm and 1.23 nm, for Ru and Au, respectively, the sign of the velocity of the upper substrate is reversed. The simulation continues until the upper and lower substrates are completely separated.  The substrate velocity used here corresponds to that in a MEMS switch operating at $\simeq 1$ GHz. The $z$-component of the force on the upper substrate is calculated during the simulation in order to determine the force-displacement relation for the contact simulation system.  We investigate the formation of defects using the  order parameter recently developed by~\citet{Ackland:2006qo}.

\section{Interatomic potentials}

The gold interatomic potential employed in the present study is an embedded atom method (EAM) type  potential developed by~\citet{Cai:1996kx}. This potential provides a reasonable description of both bulk and surface phenomena and was used previously in contact simulation experiments.~\cite{Cha2004,Song2006,Song2007,Song2007a}

We examined several potentials for ruthenium that were available in the literature.  For the present simulations, we were looking for a potential that reproduces the elastic constants and cohesive energy of Ru, and also gives stacking fault energy and surface energies that are in reasonable agreement with experiment. Stacking fault energies play an important role in plastic deformation and surface energy determines the work of adhesion - both key elements in contact formation and separation phenomena. \citet{Igarashi:1991lr} developed two different EAM (Finnis-Sinclair) potentials for Ru. The stacking fault energies were $I_{2}$= 13 $ [\rm  meV/\AA^{2}]$ in one version and $I_{2}$= 30.54 $ [\rm  meV/\AA^{2}]$ in the other.  Both of these differ significantly from the experimental value of $I_{2}$= 53.8 $ [\rm  meV/\AA^{2}]$.  Hence, this potential is unsuitable for our contact simulations. While modified embedded atom method (MEAM) potentials~\cite{Baskes:1992fk,Baskes:1994qy} have significantly more flexibility, the MEAM potentials fail to produce a stable hcp crystal, as noted previously by~\citet{Mae:2020uq}.  We confirmed this via MD simulations, which showed that application of a small strain destabilized the hexagonal close packed lattice of MEAM Ru.  \citet{Grinberg2000} developed an EAM potential for Ru, but did not include (or calculate) the surface and stacking fault energies.  Our experience shows that such properties will not be reproduced unless they (or surrogate properties) are included in the fit.  \citet{Hu:2001fj} recently proposed another Ru potential, but the calculated values of the stacking fault energy  $I_{2}$= 4.25 $ [\rm  meV/\AA^{2}]$ and the surface energy of the basal plane\ $\gamma$= 80 $ [\rm  meV/\AA^{2}]$ are far from the experimental values. 
\begin{table*}[htbp]
   \centering
 \caption{The analytical form of the Ru potential. All distances are expressed in $\rm \AA$ and energies in eV. The two numbers listed as "cutoff" indicate the range of the adjacent basis function: a function with cutoff $x - y$ should be multiplied by $H(r - x)H(y - r)$, where $H$ is the Heaviside step function. The potential functions are available in a tabulated format at http://www.ctcms.nist.gov/potentials}
   \begin{tabular}{l|l|l}
   \hline
       Function   & Value & Cutoff \\
    \hline
V(r) & $   \exp(11.393523549007-8.8769414996354 r+3.7751688938761r^2-0.99892512021332 r^3)$& 1.0-1.9 \\
&$+84.370697642971 (2.9-r)^4 -370.62682398816 (2.9-r)^5$&1.9-2.9\\
&$+655.67905839695 (2.9-r)^6-549.25562909352 (2.9-r)^7$&1.9-2.9\\
&$+177.09076987937 (2.9-r)^8$&1.9-2.9\\
&$-3.0363100686117 (5.0-r)^4-3.0909218331545 (5.0-r)^5$&1.9-5.0\\
&$-4.4211545804424 (5.0-r)^6-1.9205251191921 (5.0-r)^7$&1.9-5.0\\
&$-0.31060214601543 (5.0-r)^8$&1.9-5.0\\
&$+0.93316319033507 (6.4-r)^4-3.0761272816964 (6.4-r)^5$&1.9-6.4\\
&$+3.5391800334269 (6.4-r)^6-1.7626076559210 (6.4-r)^7$&1.9-6.4\\
&$+0.32587864688719 (6.4-r)^8$&1.9-6.4\\
    \hline
$\Phi(r)$ & $1.180972686 (4.8-r)^4$& 0-4.8\\
& $-0.7740233148 (4.8-r)^5+0.2696095863 (4.8-r)^6$& 0-4.8\\
& $-0.0471950464 (4.8-r)^7 +0.0032605667 (4.8-r)^8$ & 0-4.8\\    
    \hline   
F($\rho$) &$0.88124070590302 \rho^{0.5}$& 0-$\infty$\\
&$+6.5448285611211 \times 10^{-7} (\rho-65)^4$& 65-$\infty$\\
&$-2.5046137745775 \times10^{-6} (\rho-75) ^4$& 75-$\infty$\\
&$+2.3771143627356\times 10^{-5} (\rho-90)^4$& 90-$\infty$\\
&$-4.4929936875438\times 10^{-5} (\rho-95)^4$& 95-$\infty$\\
&$+2.3927653989776\times 10^{-5} (\rho-100)^4 $& 100-$\infty$\\    
   \end{tabular}
   \label{tabpot}
\end{table*}

Because none of the extant Ru potentials met the requirements for contact simulations, we develop a new EAM potential that reproduces the stacking fault energy and the surface energy of ruthenium with good accuracy. 
The EAM potential between the atoms is given by~\cite{Finnis:1984ca} 
\begin{equation}
U=\sum_{i=1}^{N-1}\sum_{j=i+1}^{N} V(r_{ij}) +\sum_{i=1}^{N}F(\rho_{i}) \ ,
\label{eq:eam}
\end{equation}
where the subscripts $i$ and $j$ indicate each of the N atoms in the system, $r_{ij}$ is the distance between atoms $i$ and $j$, $V(r_{ij})$ is a pairwise potential, $F(\rho_{i})$ is the embedding energy function and $\rho_{i}=\sum_{j} \Phi(r_{ij})$ is the radially-symmetric electron density at the position of atom $i$. The function $\Phi(r)$ is another pairwise potential.  Creating the new potential involved finding optimal functions for $V(r_{ij})$, $F(\rho_{i})$, and $\Phi(r_{ij})$ of Eq. (\ref{eq:eam}). The potential development procedure was described in details in Ref.~\onlinecite{Mendelev:2003kl}.   The analytical form of the functions for Ru is reported in Table~\ref{tabpot}.
\begin{table*}[htbp]
   \centering
 \caption{ Values of several physical properties of ruthenium determined using the new potential developed herein compared with values from calculations or experiments. The properties include the lattice constants $a$ and $c$ (Ref.~\onlinecite{Barrett:1966}), elastic constants ($C_{11}$, $C_{12}$, $C_{13}$, $C_{33}$, $C_{44}$,  Ref.~\onlinecite{Brandes:1992qy}), {\bf $(0001)$} surface energy $\gamma$ (Ref.~\onlinecite{Tyson:1977}), cohesive energy $E_{c}$ (Ref.~\onlinecite{Kittel:1976}), vacancy formation energy $  E_{f}^{v}$ (Ref.~\onlinecite{Igarashi:1991lr}), basal stacking fault energy $I_{2}$ (Ref.~\onlinecite{Legrand:1984}), energy difference between the hcp and fcc lattices  $\Delta E_{hcp -> fcc}$, energy difference between the hcp and bcc (body centered cubic) lattices  $\Delta E_{hcp -> bcc}$ (Ref.~\onlinecite{Saunders:1988}), and melting temperature $T_{m}$ (http://www.webelements.com). All of these properties  were used in fitting the potential, except for the melting temperature.}
  \begin{tabular}{cccccc}
    \hline
    \hline
         & Literature& T=0 K & T=3 K & T=300 K   \\
    \hline
     $a$ $[ \rm \AA]$   &2.706   & 2.705 &     2.7046   & 2.7067   \\
     $ c$ $ [\rm \AA]$ &4.282 & 4.288&     4.288   & 4.288\\
     $ c/a$ $[\rm \AA]$ &  1.582 & 1.585&     1.5854  & 1.5841 \\
     $I_{2}$ $[\rm  meV/\AA^{2}] $ & 54.6 & 53.8&       &  \\
     $ \gamma$ $ [\rm  meV/\AA^{2}]$ &  190&  189&     189 &    189 \\
     $ C_{11}$  & 563&  552$^{b}$&     514   & 476   \\
     $ C_{12} $ & 188&165$^{b}$&     148   & 152   \\
     $ C_{13} $ & 168& 170$^{b}$&     145$^{a}$  & 162$^{a}$  \\
     $ C_{33} $ & 624&611$^{b}$&     586   &  610   \\
    $  C_{44} $ & 181&198$^{b}$  &     199   & 197  \\
    
     $  E_{c}$ $ [\rm  eV] $ & -6.74 & -6.864 &     -6.864   & -6.777 \\
     $  E_{f}^{v}$ $ [\rm  eV] $& 1.85& 2.218 &       &  \\
     $  \Delta E_{hcp -> fcc}$ $ [\rm  eV] $& 0.125 & 0.152&     0.153  &  0.137   \\
      $  \Delta E_{hcp -> bcc}$ $ [\rm  eV] $ & 0.265 & 0.216&   bcc unstable  & bcc unstable     \\
      $T_{m}$ [K] & 2607  & 1792 & & &\\
   \hline
   \hline
   \end{tabular}
   \label{tabprop}
   \flushleft
    ({\it a})~Average of $C_{13}, C_{31}, C_{23}$. \\
({\it b})~Calculated using a virial expression. 
   \end{table*}

Table~\ref{tabprop} shows a comparison between several crystal properties determined from the potential and values either determined from calculations or experiments.
Note that lattice constants, surface energy $\gamma$ and basal stacking fault energy $I_{2}$ are in very good agreement with the experimental values.~\cite{Barrett:1966,Tyson:1977,Legrand:1984} We obtained larger differences between our calculated values and experimental values of the other properties. In particular, we determine the melting point of this Ru potential to be $T_m=1792\pm5$ K, using the coexistence method.~\cite{Morris:1994}  While this is low compared with the experimental value for Ru $T_m=2607$ K, this is not a problem since all simulations were performed at temperatures less than a quarter of the melting temperature. 
Overall the potential is much better suited for contact simulations than previous potentials. 
Furthermore, we performed a series of MD simulations using this potential over a wide range of temperatures in order to insure that the hcp structure is stable at finite temperatures.  These simulations were performed on a perfect crystal containing 4800 Ru atoms and using periodic boundary conditions in all three directions, where the periodic lengths were free to adjust to insure zero stress in the system.  The hcp structure was found to be stable from low temperature up to the melting point.

\section{Ruthenium Contacts}
\begin{figure}[htbp]
   \centering
   \includegraphics[width=8cm]{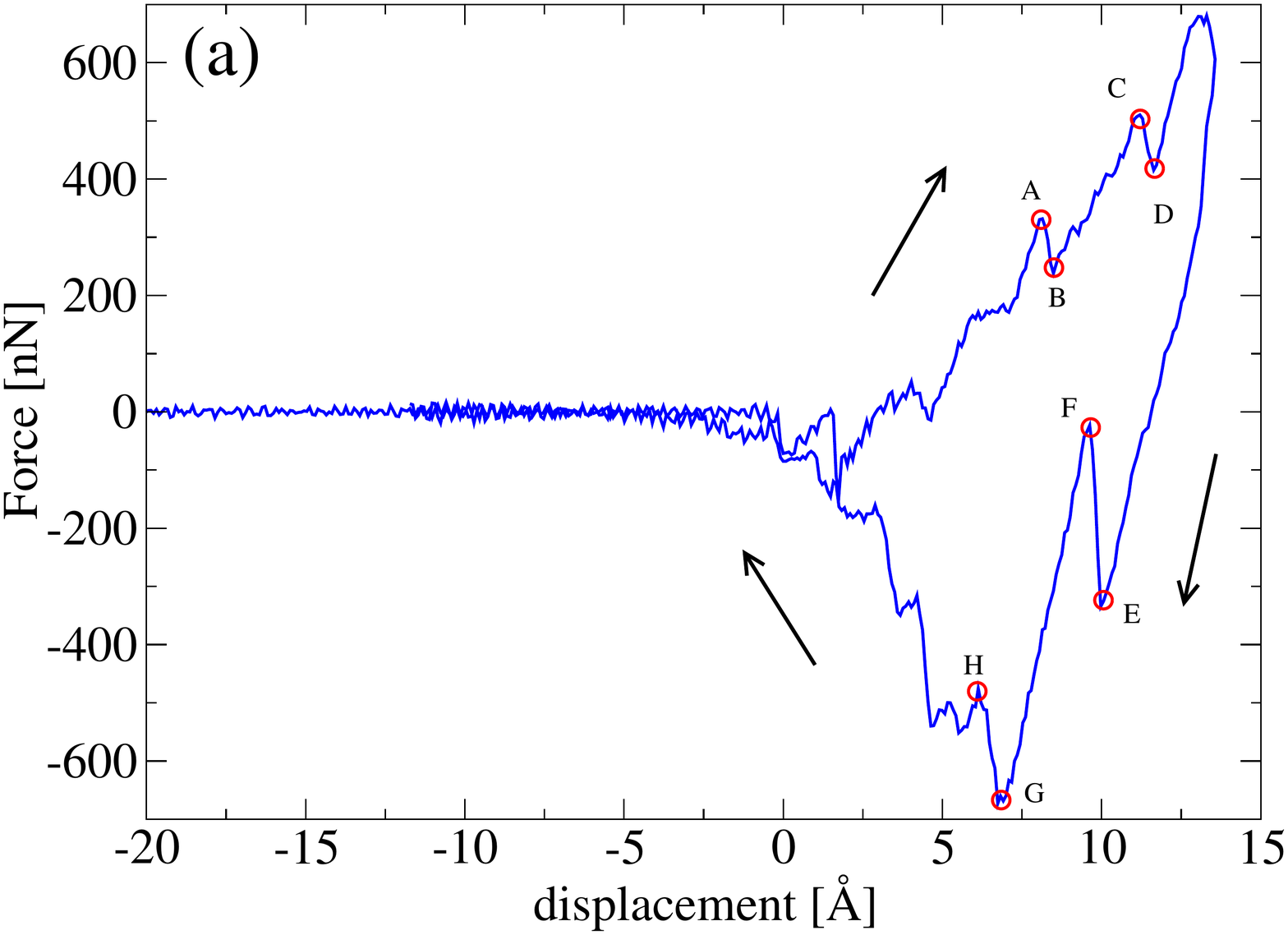} 
   \includegraphics[width=8cm]{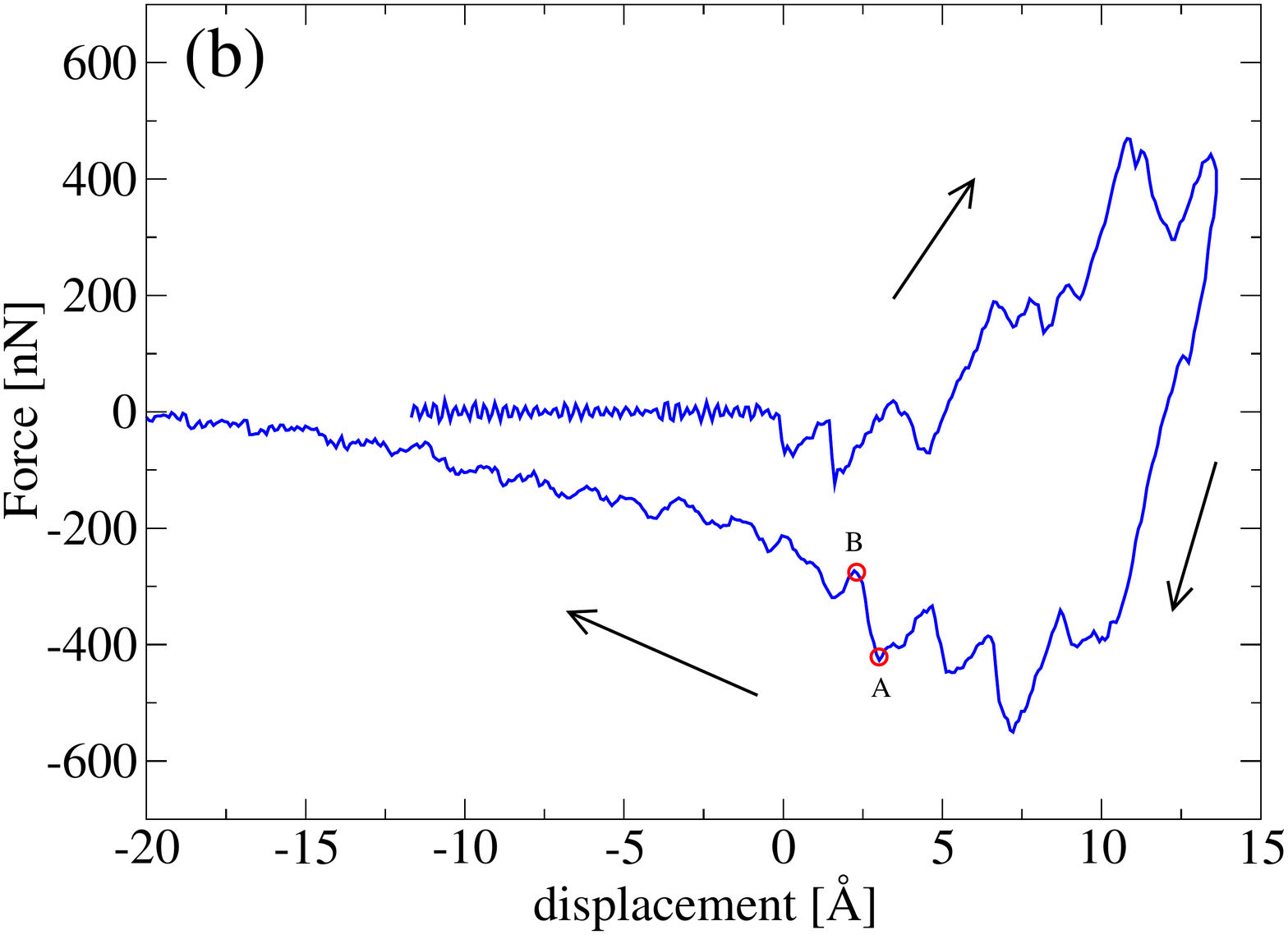} 
   \caption{(Color Online) Force-displacement diagram for Ru contacts for (a) T=300 K and  (b) T=600 K. Positive (negative) force indicates compression (tension).  The capital letter labels in the plots correspond to configurations discussed in the text. The arrows indicate the direction in which the curves are traversed.}
   \label{fig:rufd300}
\end{figure}

Figures~\ref{fig:rufd300} shows the force-displacement curves for the Ru contact simulations at  temperature T=300 K  and T=600 K. 
The two force-displacement curves have many features in common. The zero of the displacement corresponds to the first contact between the top substrate and the asperity;  at large separation (initially negative displacement), the force is zero.  As the two substrates approach each other, the force becomes negative because of the short range attractive interactions between the atoms on the opposing substrate surfaces  (this is determined by the interaction range of the interatomic potentials). This attraction elastically stretches the two materials and there is a jump-to-contact.~\cite{Smith:1989,Landman:1990,Untiedt:2007}  As the displacement increases beyond this point, the material goes into compression.  There is an approximately linear rise in the force with displacement, punctuated by  a series of relatively sharp drops. The linear increase between the drops corresponds to elastic compression. The sharp drops correspond to defect generation or annihilation events (see below).  When the sign of the velocity changes, the system begins to unload from the compressed state.  The unloading is initially characterized by linear elastic regions with a few jumps.  Eventually, the force reaches a minimum (i.e., a maximum tensile stress), following which the tensile stress slowly decreases to zero over a long displacement range.  In this region, the overall force-displacement trend is also interrupted by sharp jumps.  

\begin{figure*}[htbp]
   \centering
  \includegraphics[width=14cm]{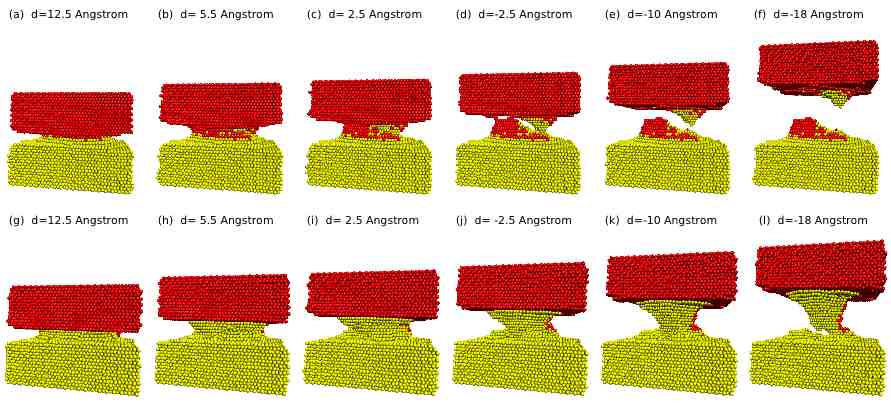} 
    \caption{(Color Online) A series of configurations corresponding to decreasing displacements  $\rm d$ in Angstroms, during the separation of the Ru contact surfaces at (a)-(f) $300$ K and (g)-(l) $600$ K. The atoms are all Ru, but are colored red or yellow to indicate that they were initially part of the upper or lower substrates, respectively. }
   \label{fig:snruseq}
\end{figure*}

The two substrates separate at larger (negative) displacement than that at which the initial contact occurred (i.e., zero displacement).  The magnitude of the displacement necessary to separate the two substrates is much larger in the high temperature simulation than for the low temperature one, suggesting that the material is much more ductile (larger strain to failure) at high temperature than at low temperature.  This interpretation  is easily confirmed by reference to the morphologies observed during deformation at $300$ K and $600$ K shown in Fig.~\ref{fig:snruseq}(a)-(f) and Fig.~\ref{fig:snruseq}(g)-(l), respectively.
At T=300 K (Fig. \ref{fig:snruseq}(a)-(f) ) we observe a fracture-like rupture of the bridge of atoms connecting the two substrates. On the other hand, at T=600 K (Fig.~\ref{fig:snruseq}(g)-(l)) we observe the formation and plastic stretching of a relatively long neck between the two substrates.  

\begin{figure}[htbp]
   \centering
   \includegraphics[width=8cm]{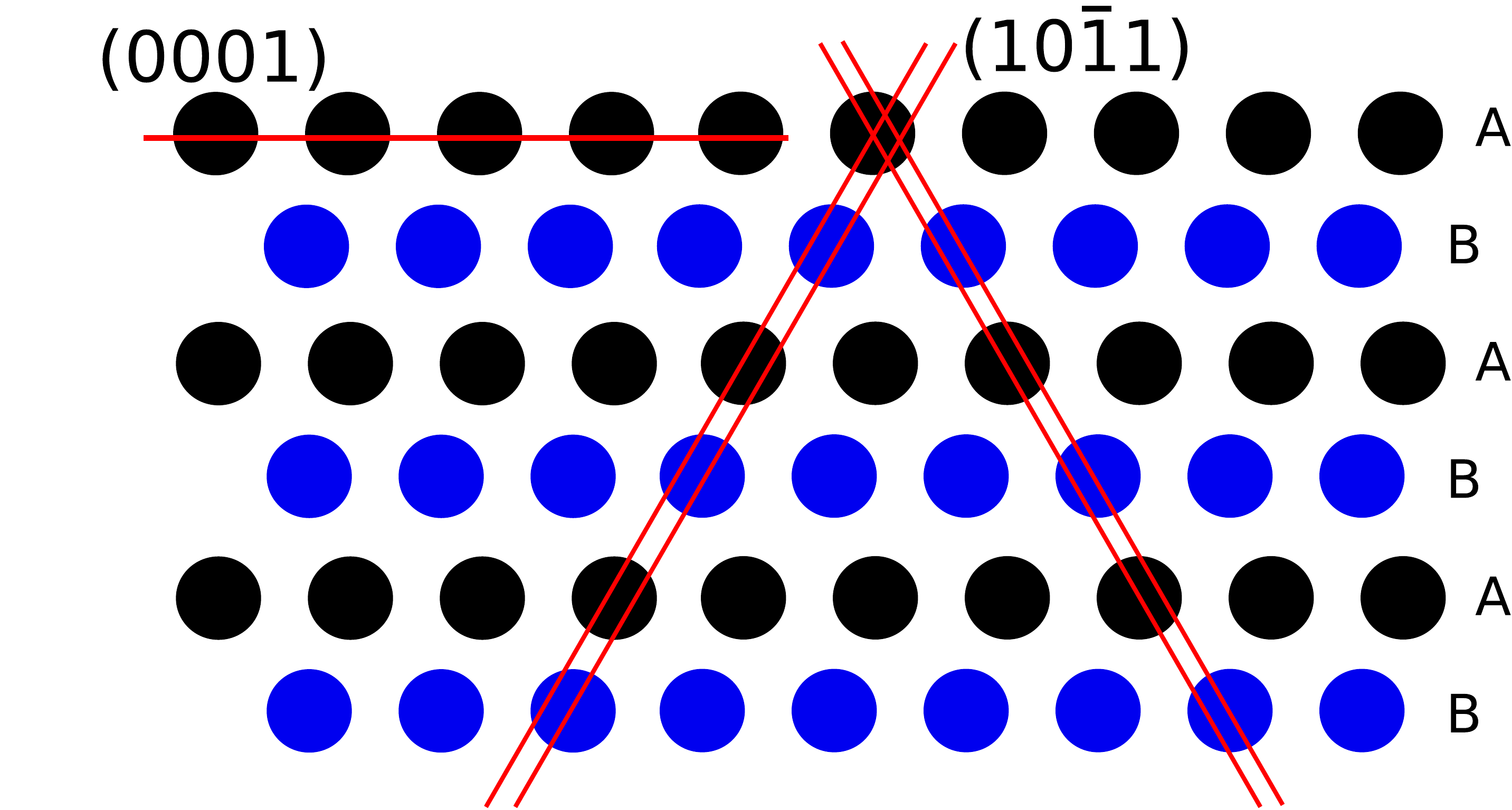} 
   \caption{(Color Online) Schematic representation of the $[ 1 1 \bar 2 0]$ projection of the hcp structure. The single and double red lines indicate a $(0001)$ basal plane and a pair of ${10\bar11}$  pyramidal planes, respectively. The letters, A and B, refer to stacking sequence of the basal planes.  The coloring is imply to distinguish between these planes.}
   \label{fig:sc}
\end{figure}

Before we begin the examination of the atomic configuration and how it evolves during contact and contact separation, we first remind the reader of the hcp structure, as viewed in the projection employed in many of the images in the remainder of the paper.  Figure~\ref{fig:sc} shows the hcp structure viewed along the  $[ 1 1 \bar 2 0]$ direction (i.e., the $( 1 1 \bar 2 0)$ plane). In this view, basal  $( 0 0 0 1)$ and pyramidal  $( 1 0 \bar1 1)$ planes along which dislocations can glide are visible. We will also show views along the $[ 0 0 0 1]$ directions; i.e., the basal planes. The $ [0 0 0 1 ]$ projection shows dislocations thanks to the atom mismatch in different planes.

\begin{figure}[htbp]
   \centering
   \includegraphics[height=7cm]{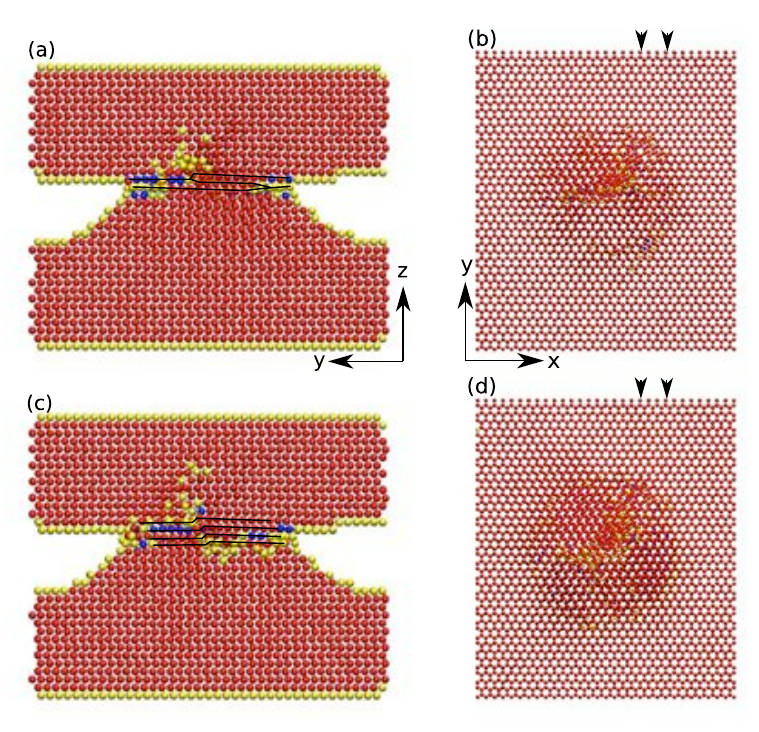}  
   \caption{(Color Online) Atomistic configuration during the compression of a Ru asperity (lower substrate) against a flat Ru surface at T=300 K.  Red atoms are in an hcp environment, blue atoms within an fcc environment and yellow atoms are in an environment of some other, unknown symmetry.  (a) and (b) correspond to point A in Fig.~\ref{fig:rufd300}(a).  (a) Lateral view of a thin slab of four $( 1 1 \bar 2 0)$ planes, while (b) the top view of the contacts projected on the basal $ (0 0 0 1)$ plane at point A  in Fig.~\ref{fig:rufd300}(a).  (c)  Lateral view of a thin slab of 4 $( 1 1 \bar 2 0)$ planes at point B  in Fig.~\ref{fig:rufd300}(a). (d)  The top view of the contacts projected on the basal $ (0 0 0 1)$ plane at point B in Fig.~\ref{fig:rufd300}(a). Lines in (a),(c) are guides to the eye.  The arrows in (b),(d) indicate the position of the slab section reported in (a),(c).}
   \label{fig:snru1}
\end{figure}

To determine the microscopic origin of the force drops in the force-displacement curve, we examine the structure for the presence, formation and annihilation of defects.  Based on earlier contact simulation observations~\cite{Cha2004,Song:2008} and the fact that dislocations carry plastic deformation, we focus upon dislocations.  The local atomic environment around dislocations and the stacking faults that they can create differ from those in the perfect crystal.  Therefore, we employ the order parameter developed by \citet{Ackland:2006qo} to describe the environment around individual atoms since it is able to distinguish between different local configurations (including those of different crystal structures). The presence of local structures of different or unknown symmetry within the metal indicates the presence of a defect.  In the images discussed below, we color code the atoms according to this symmetry parameter, coloring atoms red in an hcp environment, blue in an fcc environment and yellow, otherwise.  A stacking fault along the $(0 0 01)$ planes of the hcp structure will appear as a plane of blue atoms in the red matrix.  Dislocation cores and stacking faults in other planes (such as the pyramidal planes) will appear yellow.
\begin{figure}[htbp]
   \centering
   \includegraphics[height=7cm]{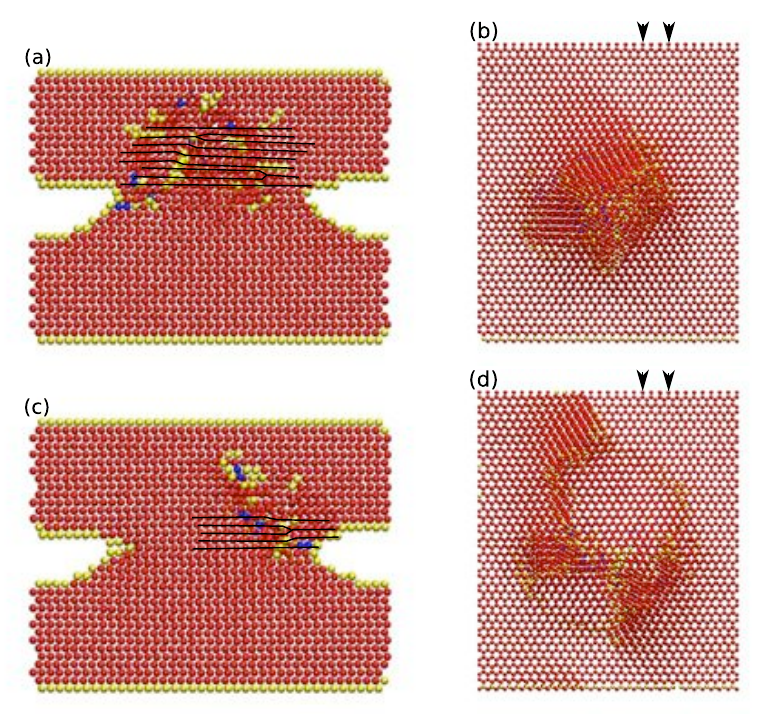}    
   \caption{(Color Online) Same as figure~\ref{fig:snru1}, but for point  C  (a),(b) and point D (c),(d) in figure~\ref{fig:rufd300}(a). }
   \label{fig:snru2}
\end{figure}

\begin{figure}[htbp]
   \centering
   \includegraphics[height=7cm]{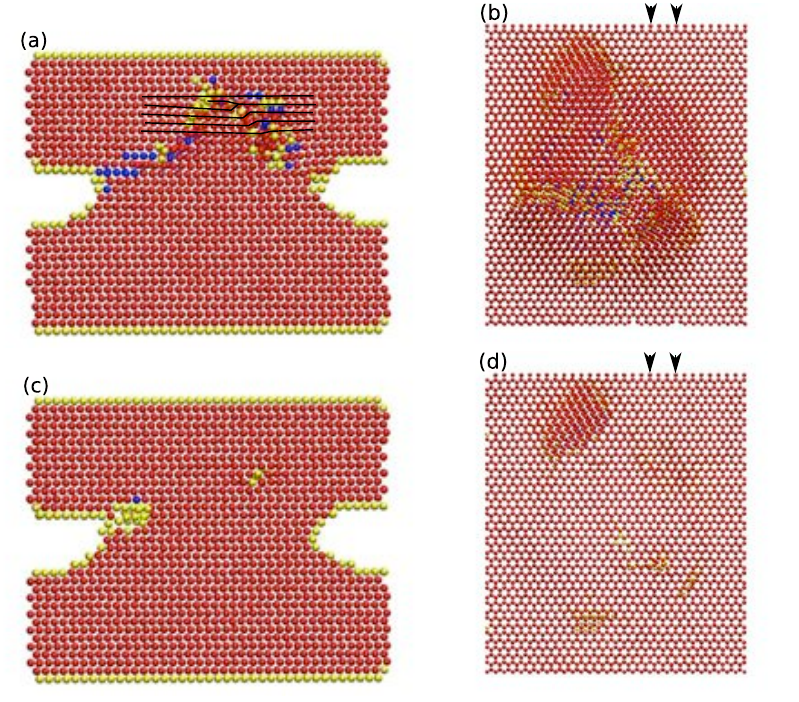}   
   \caption{(Color Online) Same as Fig.~\ref{fig:snru1}, but during unloading, at point  E  (a),(b) and point F (c),(d) in Fig.~\ref{fig:rufd300}(a).}
   \label{fig:snru3}
\end{figure}

Figure~\ref{fig:snru1}(a) shows the atomic configuration in a four $( 1 1 \bar 2 0)$ atomic plane thick slap  which cut the system near its center in a direction orthogonal to the substrate. This configuration corresponds to the point A in the force-displacement curve of Fig.~\ref{fig:rufd300}(a). Figure~\ref{fig:snru1}(b) shows the projection of the system at the same point on a $( 0 0 0 1)$ plane parallel to the substrate.  When stress is applied in the $ [0 0 0 1 ]$ direction, we find that dislocations nucleate in the corners of the asperity/substrate contact, as seen in Fig.~\ref{fig:snru1} (a) and (b). Careful examination of the dislocation produced shows that the active slip system here is  $ \{1 0\bar 1 1  \} \langle 1 \bar 2 1 3  \rangle$ indicating the presence of dislocations with Burgers vector of the$ \frac{1}{3} \langle \bar 1 \bar 1 2 3  \rangle$ type.  These dislocations  dissociate on the pyramidal  and basal plane.~\cite{Minonishi:1981,Numakura:1990,Numakura:1990a}  The small force drops in the force-displacement curve are associated with plastic events in the system, like the formation or annihilation of these dislocations. 

Figure~\ref{fig:snru1}(c) and~\ref{fig:snru1}(d) show the configuration corresponding to the point B in Fig.~\ref{fig:rufd300}(a). The large force drop at B occurs due to the disappearance of half of a $( 0 0 0 1)$  plane at the interface - originally as seen in Figs.~\ref{fig:snru1}(a),(c). The edge dislocation in the basal plane disappears at point B [Fig.~\ref{fig:snru1}(c),(d)], leaving behind an asperity with one less $( 0 0 0 1)$ atomic plane.  We observe that, despite the disappearance of this plane,  defects  accumulate within the upper substrate. Careful examination of Fig.~\ref{fig:snru1}(d) shows that at least four lines of yellow atoms (dislocations) on different  $\{ 1 0 \bar 1 1\}$ planes are visible.  This is not surprising since  several   $\{ 1 0 \bar 1 1\}$ planes have the same orientation with respect to the loading axis.  This results in an asperity that is broadening symmetrically during loading.  The large force drops during loading are associated with the disappearance of $\{ 0 0 0 1 \}$ planes from the asperity. (Similarly, large force jumps during contact separation correlate with the creation of new $\{ 0 0 0 1 \}$ atomic planes in the asperity.)

During loading, defects continue to accumulate in the proximity of the asperity as shown in Figs.~\ref{fig:snru2}(a),(b)). This state corresponds to point C  in Fig.~\ref{fig:rufd300}(a).
Between point C and D, a force drop occurs, corresponding to the decrease of one atomic plane in the asperity as shown in Figs. \ref{fig:snru2}(a),(c).  This takes place through a series of slip events that have the effect of expelling the defects outside of the asperity region. (Figs. \ref{fig:snru2}(c),(d)).
 \begin{figure}[htbp]
   \centering
   \includegraphics[height=7cm]{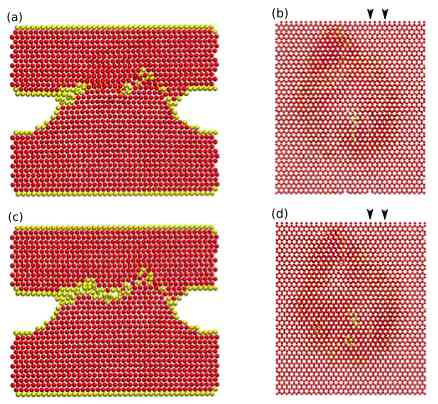}    
   \caption{(Color Online) Same as Fig.~\ref{fig:snru1}, but during unloading at point  G  (a),(b) and point H (c),(d) in Fig.~\ref{fig:rufd300}(a).}
   \label{fig:snru4}
\end{figure}

\begin{figure}[htbp]
   \centering
   \includegraphics[height=7cm]{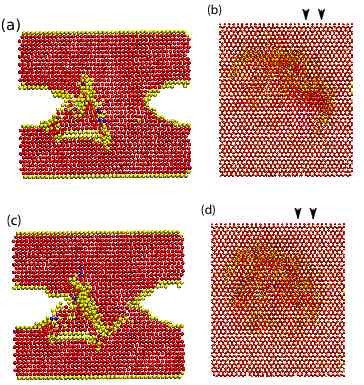}    
   \caption{(Color Online) Same as Fig.~\ref{fig:snru1}, but for temperature T=600 K during unloading.  Points  A and B  in Fig.~\ref{fig:rufd300}(b) are shown in (a),(b) and (c),(d), respectively.}
   \label{fig:snru6}
\end{figure}

The processes shown in Figs.~\ref{fig:snru1} and~\ref{fig:snru2} repeat until the contacts begin to separate.
In the early stages of the unloading process, the force decreases linearly as the top substrate is retracted.  At T=300 K,  when the unloading process is started, dislocations are still present in the system. These defects survive the early stages of the unloading and at point E in Fig.~\ref{fig:rufd300}(a), the  dislocations are extended, as shown in Fig.~\ref{fig:snru3}(a),(b). At point F in Fig.~\ref{fig:rufd300}(a), the defects have almost completely disappeared through a series of slip events in the pyramidal planes (Fig.~\ref{fig:snru3}(b),(c)). During this process, a new plane is created inside the bridge of atoms, and a large force jump is observed in the force-displcement curve.
After this first event, the elastic response continues until the slope of the force-displacement curve changes.
At this point, we observe both crack propagation along with slip. Both cracks and dislocations are nucleated at the asperity/substrate contact (Fig.~\ref{fig:snru4}). 
In particular, from the top view, we observe how plastic events occur mainly near the outer surface of  the asperity, while in the center we observe fracture, as seen from the near perfection in the cross-sections in Fig.~\ref{fig:snru4}(b),(d).
These fracture events correlate with small jumps in the force-displacement curve.

When the contact simulation is performed at T=600 K, we observe the same general features as seen at the lower temperature, T=300 K, during the loading portion of the cycle. The same slip events occur, but with less accumulation of defects in the substrate.  This is most likely the result of the larger  dislocation mobility at elevated temperature.
On the other hand, the unloading process is very different. Although the same slip system, $ \{1 0\bar 1 1  \}$$ \langle 1 \bar 2 1 3  \rangle$, is active at T=600K, the deformation at T=600K is not accompanied by fracture.  Dislocations are nucleated at the corner of the bridge/substrate contact. 
The bridge elongates by  repeated slips along these planes and the rearrangement of the atoms to form new atom layers (plastic accommodation of the tensile strain in the $z$-direction).
An example of the slip process is shown in Fig.~\ref{fig:snru6}.

\section{Gold Contacts}
In order to compare the behavior of Ru to that of the well studied case of Au, we repeat the contact simulations for Au at approximately the same homologous temperature $T_{h}=T/T_{m}$, where $T_{m}$ is the melting temperature. Since, the melting temperature of Ru is $T_{m}$=2607 K and  Au melts at $T_{m}$=1337 K  we now analyze Au at  T=150 K and T =300 K.  To make the conditions as similar as possible between Au and Ru we apply the stress in the $\{111\}$ direction of Au, that is perpendicular to the hexagonal plane of the fcc lattice which is akin to the basal $\{0001\}$ plane of hcp Ru. Since the effect of orientation was studied in detail before~\cite{Song:2008}, we only briefly analyze the morphology and force-displacement curve of $\{111\}$-oriented Au here. 
\begin{figure}[htbp]
   \centering
   \includegraphics[width=4cm]{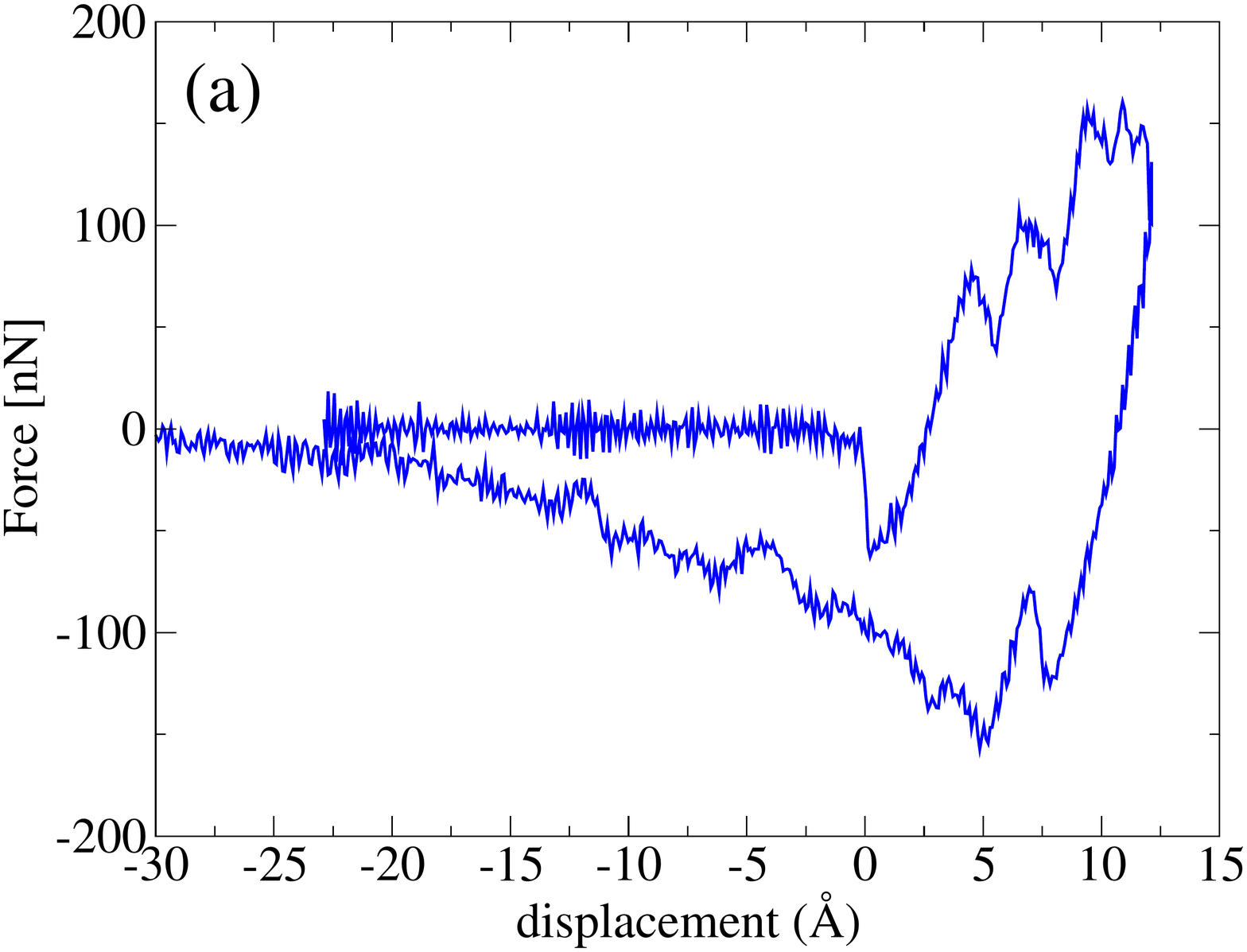} 
    \includegraphics[width=4cm]{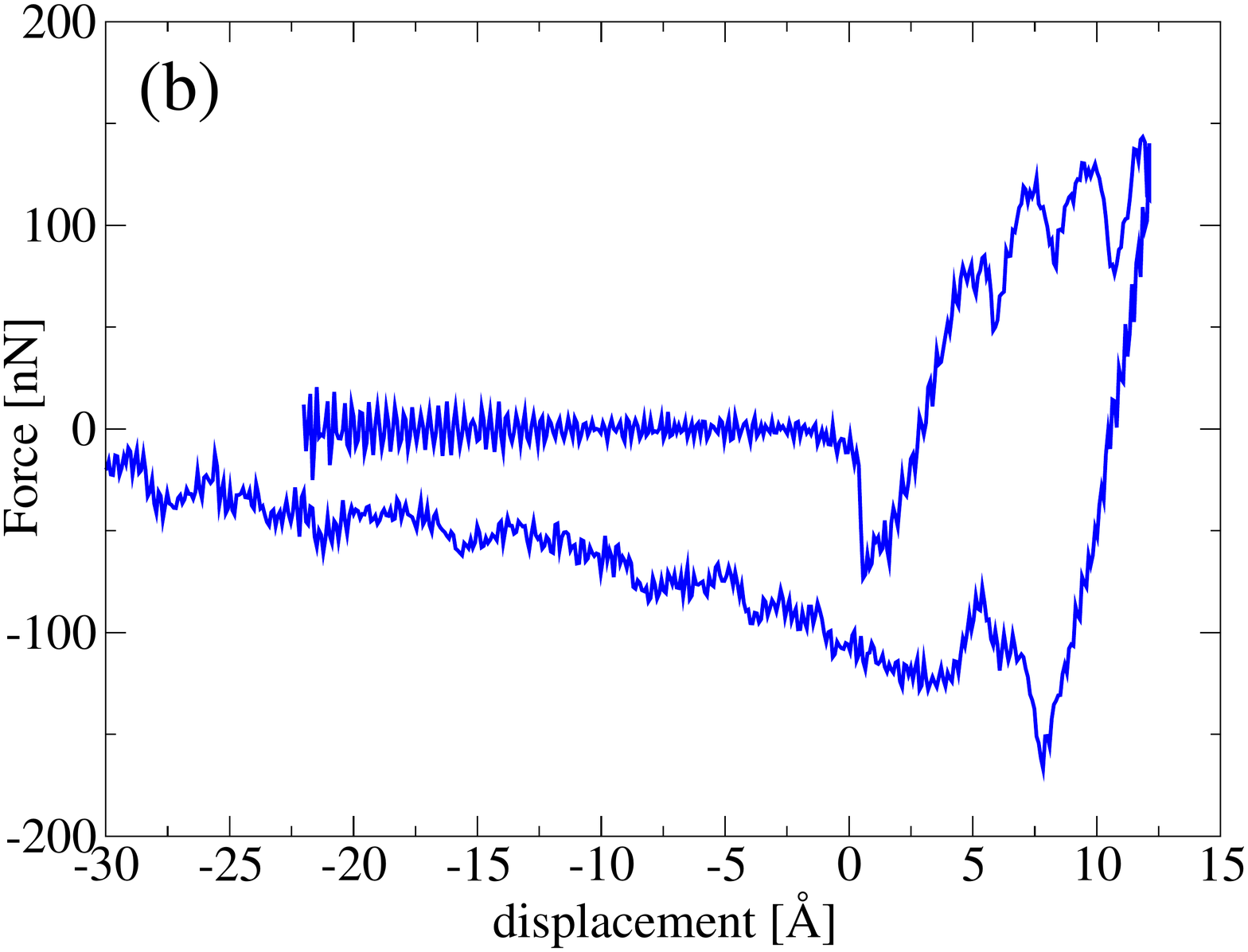} 
   \caption{(Color Online) Force-displacement diagram for Au contacts at (a) T=300 K and (b) T=150 K.}
   \label{fig:aufd300}
\end{figure}
\begin{figure}[htbp]
   \centering
   \includegraphics[width=7cm]{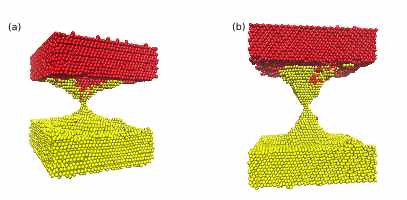} 
   \caption{(Color Online) Simulation images of Au contact separation at (a) T=300 K and (b) T=150 K. Red (yellow) spheres indicate gold atoms that at the beginning of the simulation are in the top (bottom) substrate. }
   \label{fig:snau}
\end{figure}

The force displacement curves for $\{111\}$-oriented Au at T=150K and T=300K are shown in Figs.~\ref{fig:aufd300}.  The long tails during the separation process indicate that ductile separation (without crack propagation) occurs at both temperatures.
This is confirmed by observation of the morphology in Fig.~\ref{fig:snau}. A long, thin, connective neck forms and elongates as the two substrates are pulled apart. The neck breaks when it thins to a diameter of 1-2 atoms. This kind of separation process is responsible for the observed long tail in the force-displacement unloading curve for Au. 
The amount of material transfered between the two surfaces is about the same at the two different temperatures. Since gold has  an fcc lattice, partial dislocations tend to form and move on $ \{1 1 1  \}$ planes. Indeed, we find that with the stress oriented in the $ [1 1 1  ]$ direction, the microscopic mechanism for plastic deformation is the same as was previously observed~\cite{Cha2004,Song:2008} with the stress in the $ [1 0 0  ]$ direction.

\section{Discussion}
\begin{figure}[htbp]
   \centering
    \includegraphics[width=6cm]{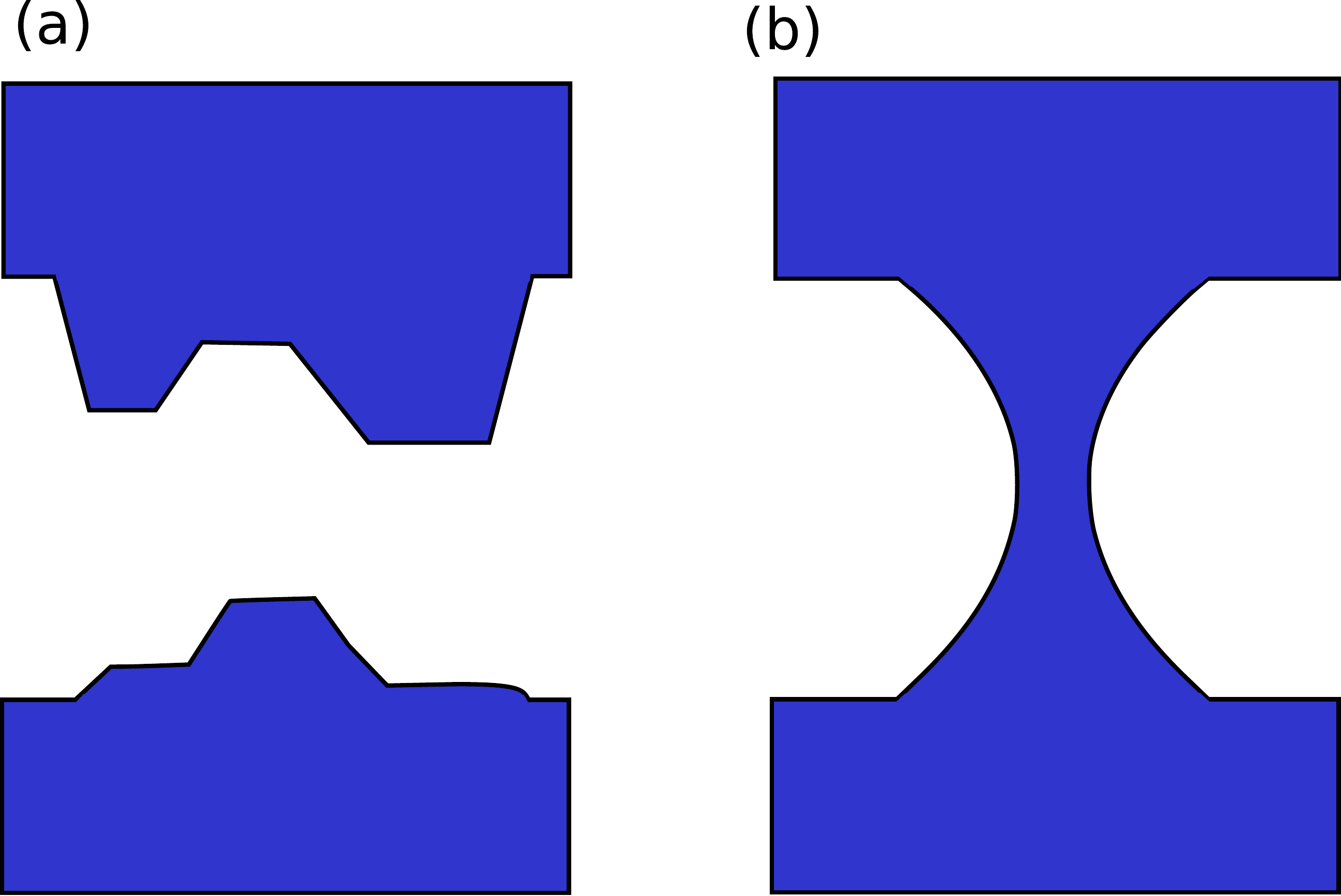} 
   \caption{(Color Online) Schematic representaton of the (a) brittle (several intersecting cracks) and (b) ductile separation scenarios. }
   \label{fig:bd}
\end{figure}

\begin{figure}[htbp]
   \centering
  \includegraphics[width=3cm]{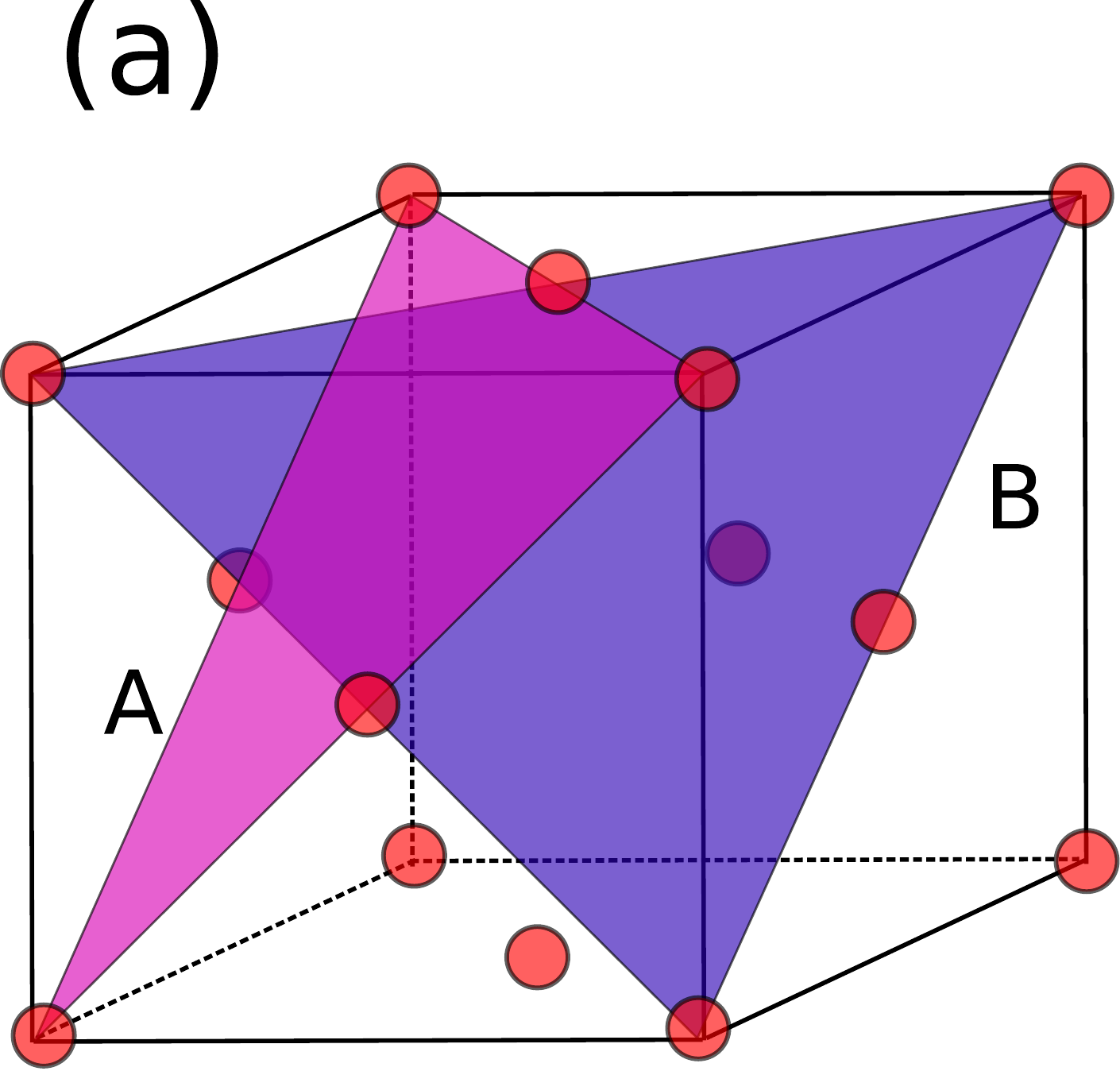} 
   \includegraphics[width=3cm]{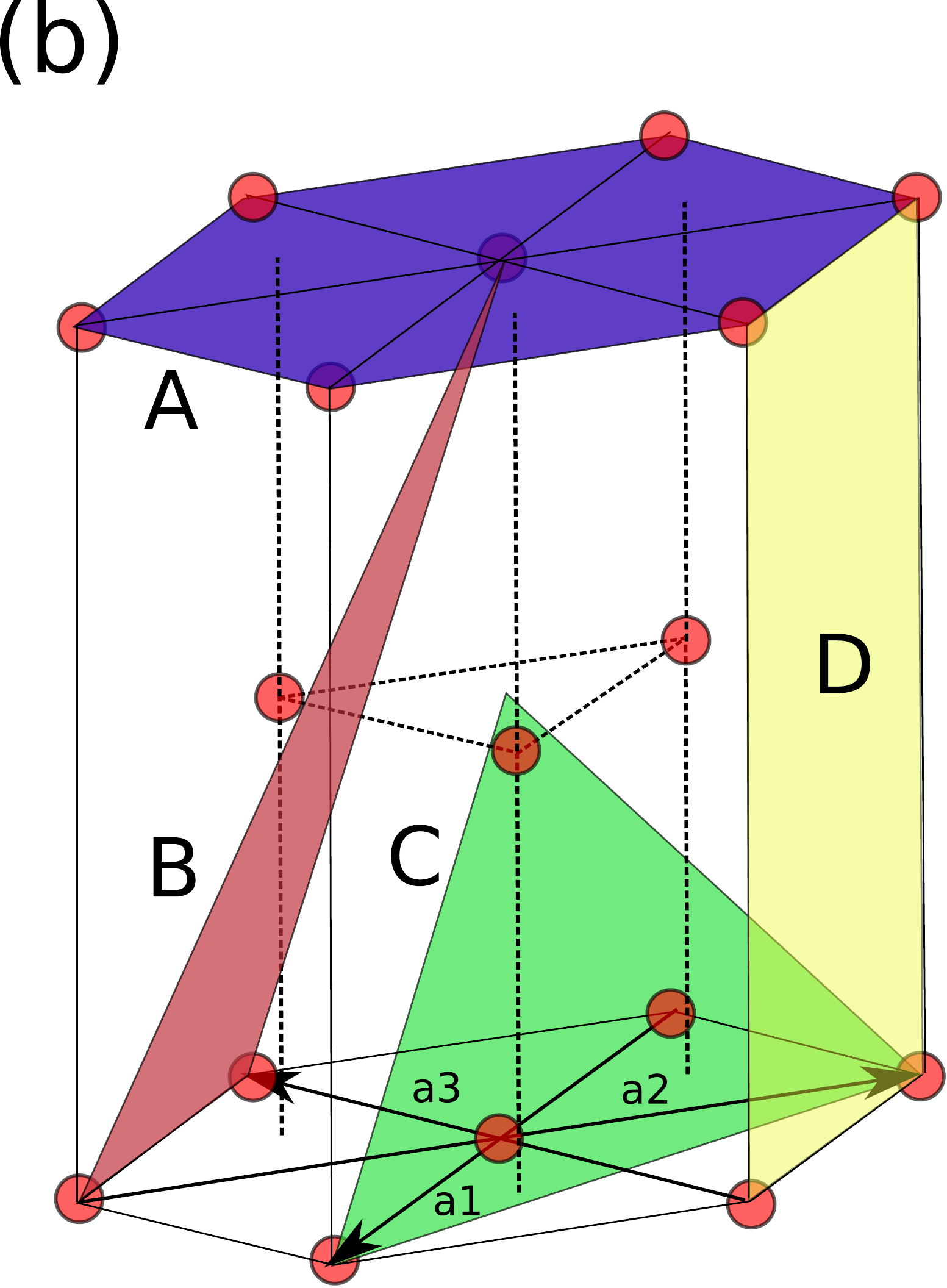} 
   \caption{(Color Online) (a) Two $\{111\}$ slip planes in the fcc (planes A and B) lattice.  
   (b) Several different possible slip planes in the hcp lattice;  A is the  $(0001)$ slip plane, B is the $(0 \bar 1 1 1)$  slip plane, C is the $ (1 1 \bar 2 2) $ slip plane, and D is the $ (0 1\bar 1 0  )$ slip plane. }
   \label{fig:geo}
\end{figure}

Operationally, we consider a material to be  brittle when separation occurs without plasticity and ductile when there is considerable plasticity and fracture/bond breaking is very limited.  Figures~\ref{fig:bd} (a) and (b) show  schematic representations of brittle and ductile separation, respectively.  Clearly, different degrees of ductility are possible, depending on the number of activate slip systems and the dislocation mobility at the test temperature. Materials that are macroscopically brittle may undergo some plasticity on the microscopic level.~\cite{Pradeep:2007}  In bulk metals, plasticity is normally controlled by the motion of dislocations.  However, in nanoscale metals, plasticity is often controlled by the nucleation of dislocations, rather than by their motion. Gold typical exhibits ductile behavior,~\cite{Kuipers:1993,Kizuka:1998,Sorensen:1998} with necking at large strains at both T=150 K and T=300 K.  Our simulations show that while Ru is ductile at T=600 K, it is more brittle at T=300 K, separating by a combination of fracture and plasticity.  The difference between the ductile/brittle behavior of Au and Ru contact has also been seen in finite element method calculations.~\cite{Du:2007ud}  There are several contributions to the difference in behavior of  Au and Ru.  Since Au is fcc and deforms by dislocation motion of the $ \{1 1 1  \}$$ \langle  0 1 1  \rangle$ type, there are many slip systems available to accommodate strain along any axis (see Fig.~\ref{fig:geo}(a)). 
On the other hand Ru has an hcp lattice, and although a number of different slip systems have been observed (Fig.~\ref{fig:geo}(b)),~\cite{Bacon:2002,Yoo:2002,Bere:2005} some of these are difficult to activate.  As a result,  the total number of slip systems that participate in deformation can be activated at low stress is more limited than in fcc. We have also performed a series of simulations in which a Ru contact was loaded along the 
$ [1 0 \bar 1 2 ]$ direction such that both the basal and pyramidal plane are active slip systems.  Here too we observed the Ru contact to separate in a brittle way at room temperature and with more ductility at T=600 K. 
The surface energy of Ru is approximately 3 times larger than in Au and the stacking fault energy of Ru is roughly 16 times higher than in Au.~\cite{Baskes:1992fk}  Increasing the stacking fault energy makes slip more difficult, while increasing the surface energy (and work of adhesion) makes fracture more difficult.  
However, since the increase in stacking fault energy in Ru comparatively to Au is much greater then the increase in surface energy, the change from Au to Ru leads to an increased tendency for fracture  in Ru than in Au.  We speculate the increased resistance of Ru to plastic deformation during contact may contribute to its increased resistance to damage during repetitive contact (i.e., the morphology changes little on cycling in Ru, unlike Au). 

\section{Conclusions}

We have investigated the nano-contact mechanics of ruthenium using molecular dynamics simulations. We first developed a new EAM potential for Ru that reproduces the experimental elastic constants, surface, and stacking fault energies.  The simulations were performed in a system composed of two substrates, one with a flat surface and the other with a parallel flat surface that has a single, roughly hemispherical, asperity. During loading, the top flat substrate moves toward the lower substrate  at a constant velocity. At a preset displacement, the sign of the velocity is switched and the two surfaces pull apart, until the contact is broken.  During the simulation, the force in the system, the atomistic configuration, and the local order parameter are monitored. In Ru at  T=300 K  and T=600 K, we observed plastic events (dislocation formation, motion, and annihilation/escape) during the loading process.  This deformation was localized primarily as slip on pyramidal planes. When the two substrates are pulled apart, we observed  brittle contact separation at T=300 K.  This was characterized by crack formation and bond breaking and by a short tail in the force-displacement curve past the maximum tensile force. On the other hand, at T=600 K, the separation was much  more ductile and  was characterized by slip on pyramidal planes and no cracks. A small bridge of atoms  between the top and the bottom substrates forms in the final stages of the separation at T=600 K.  The force-displacement curves showed considerable plastic deformation following the peak tensile force - associated with the necking that led to neck formation.
We repeated nearly identical contact simulation in  Au at the same homologous temperatures, for the sake of comparison. 
At both T=150 K and T=300 K, Au is strongly ductile with a long tail in the force-displacement curve and contact separation occurs with the formation of a neck. In Au, the plastic deformation corresponds to slip along the $\{111\}$ planes. 
Our findings provide a partial explanation of why Ru contacts are more reliable than Au contacts.  That is, Ru undergoes much less plastic deformation and morphology change than does Au under the same conditions. To be a good contact, the material must have a high electrical conductance.  Unfortunately, the conductance of Ru and its failure are sensitive to oxygen and carbon contamination.~\cite{Tringe:2001} In real Ru contacts, the surfaces are typically covered with layers of ruthenium oxide that can change the adhesion and stacking fault energies.
Work is in progress to develop a model to study the effect of ruthenium oxide on the mechanical properties of nanocontacts. 

\acknowledgments 
AF thanks Jun Song for discussions.
This work was supported by DARPA under its S\&T Fundamentals program.
Work at the Ames Laboratory was supported by the Department of Energy, Office of Basic Energy Sciences, under Contract No. DE-AC02-07CH11358.


\end{document}